\documentclass{aa} 
\usepackage{epsf}
\usepackage{graphicx}

\newcommand{\msun}{$M_{\odot}$}

\newcommand{\rsun}{$R_{\odot}$}

\newcommand{\Mbol}{$M_{\rm bol}$}
\newcommand{\Mbols}{$M_{\rm bol,\odot}$}
\newcommand{\ergs}{erg\,cm$^{-2}$s$^{-1}$}

\newcommand{\teff}{$T_{\rm eff}$}
\newcommand{\mv}{$M_{\rm V}$}

\newcommand{\mk}{$M_{\rm K}$}

\newcommand{\sk}{$S_{\rm K}$}

\newcommand{\sv}{$S_{\rm V}$}

\newcommand{\mlam}{$m_{\lambda}$}
\newcommand{\Mlam}{$M_{\lambda}$}
\newcommand{\vrj}{$V-R_{\rm J}$}
\newcommand{\ik}{$I_{\rm c}-K$}
\newcommand{\vic}{$V-I_{\rm c}$}
\newcommand{\vk}{\mbox{$V-K$}}

\newcommand{\od}{$\bigcirc \hspace*{-1.65ex}\bullet$\,}
\newcommand{\pssym}[1]{\includegraphics[width=3mm]{#1}}
\begin{document}
\thesaurus{05(02.01.1, 11.01.2, 11.14.1, 11.17.3, 11.19.1)}
\title{Barnes--Evans relations for late--type giants and dwarfs} 
\author{K. Beuermann
\inst{1}
\and I. Baraffe
\inst{2}
\and P. Hauschildt
\inst{3}}
\offprints{beuermann@uni-sw.gwdg.de}
\institute{Universit\"ats-Sternwarte, Geismarlandstr. 11, D-37083
G\"ottingen, Germany \and C.R.A.L. (UMR 5574 CNRS), Ecole Normale
Sup\'{e}rieure de Lyon, F-69364 Lyon Cedex 0.7, France \and Department
of Physics and Astronomy and Center for Simulational Physics,
University of Georgia, Athens, GA\,30602-2451, USA} 
\date{Received March 23, 1999 / Accepted June 21 1999}
\authorrunning{K. Beuermann et al.}
\titlerunning{Barnes-Evans relations}
\maketitle
\begin{abstract}
The visual surface brightness of K/M giants and dwarfs with near-solar
metallicity differ slightly in agreement with the gravity effects
predicted by recent theoretical models. We show that M-dwarfs display
also a metallicity dependence of the surface brightness in the
infrared $K$-band in agreement with theory. Based on these results, we
present improved Barnes-Evans type relations and estimate the radii of
60 single or presumed M and K-dwarfs.

\end{abstract} 
\section{Introduction} 

Barnes \&\ Evans (1976) showed that a tight relation exists between
the visual surface brightness and the colour of giants. Such a
relation allows to determine the angular diameter and, if the distance
is known, the radius of a star from photometric data alone (Lacy 1977,
Dumm \&\ Schild 1998). Lacy assumed that the Barnes-Evans relation
derived for giants holds also for dwarfs, but this has never been
actually proved.

For giants, additional measurements of the angular diameters have
become available in recent years (Dumm \&\ Schild 1998, Dyck et
al. 1998, and references therein). Angular diameters of dwarfs can so
far not be measured directly, but can be derived from bolometric
fluxes and temperatures or, more accurately, from flux scaling of
model atmospheres to the low-resolution optical/IR overall spectral
energy distributions.  Leggett et al. (1996, hereafter L96) have
applied this method to 16 M-dwarfs using the advanced {\it NextGen}
M-dwarf model atmospheres of Hauschildt et al. (1999). There is
excellent agreement between the L96 radii and those predicted by
recent stellar models (Baraffe et al. 1998, hereafter BCAH98; see the
comparisons made in L96 and in Beuermann et al. 1998, henceforth Paper
I). This convergence of theory and observation is generally regarded
as a breakthrough and an important step towards a temperature and
radius scale of stars on the lower main sequence, although there is
still some concern about the remaining differences (e.g. Clemens et
al. 1998).

In this paper, we use the results of L96 to derive the surface
brightness of M-dwarfs.  We then show that the visual surface
brightness vs. Cousins \vic{} relationships for M-dwarfs differs from
that of M-giants in a colour-dependent way and find a close agreement
between observationally determined and theoretically predicted gravity
dependencies (BCAH98, Alibert et al. 1999, Hauschildt et al. in
preparation). We, furthermore, show that M-dwarfs display a
metallicity dependence of the surface brightness in the infrared
$K$-band which also agrees with that predicted.  The good agreement
between theory and observation increases our confidence in the derived
Barnes-Evans relations and the implied radius scale of M-dwarfs.

\section{The Barnes-Evans relations for giants and dwarfs}

Barnes \&\ Evans (1976) defined the surface brightness as\footnote{The
numerical constant in Eq. (1), 4.2211, equals 1 + 0.1\,\Mbols{} +
0.25\,log\,(4\,$f_{\odot}/ \sigma$), where \Mbols{}= 4.75 is the
bolometic magnitude of the Sun, $f_{\odot} = 1.368~10^6$\,\ergs{} is
the solar constant, and $\sigma$ is the Stefan-Boltzmann constant.}
\begin{equation}
F_{\lambda}  =  -0.1 m_{\lambda} -0.5\,{\rm log}\,\phi + 4.2211  
\end{equation}
where \mlam{} is the apparent magnitude at wavelength $\lambda$ and
$\phi$ is the angular diameter of the star in mas.  With the absolute
magnitude \Mlam{} and the radius $R$ of the star, Eq. (1) can be cast
into the form (e.g. Bailey 1981)
\begin{equation}
S_{\lambda}  =  M_{\lambda} + 5\,{\rm log}(R/R_{\odot}) =
-10\,F_{\lambda} + 42.368.
\end{equation}

\begin{figure*}[t]
\includegraphics[width=12cm]{beuerman.f1}
\hfill
\raisebox{11mm}{
\begin{minipage}[b]{52mm}
\caption[]{\label{mbol_radius} Visual surface brightness \sv{}
vs. Cousins \vic{} for dwarfs and giants. Dwarf data are for eight YD
field stars from L96 (\od), the mean of the two components of the
eclipsing binary YY Gem (\raisebox{-0.5ex}{\pssym{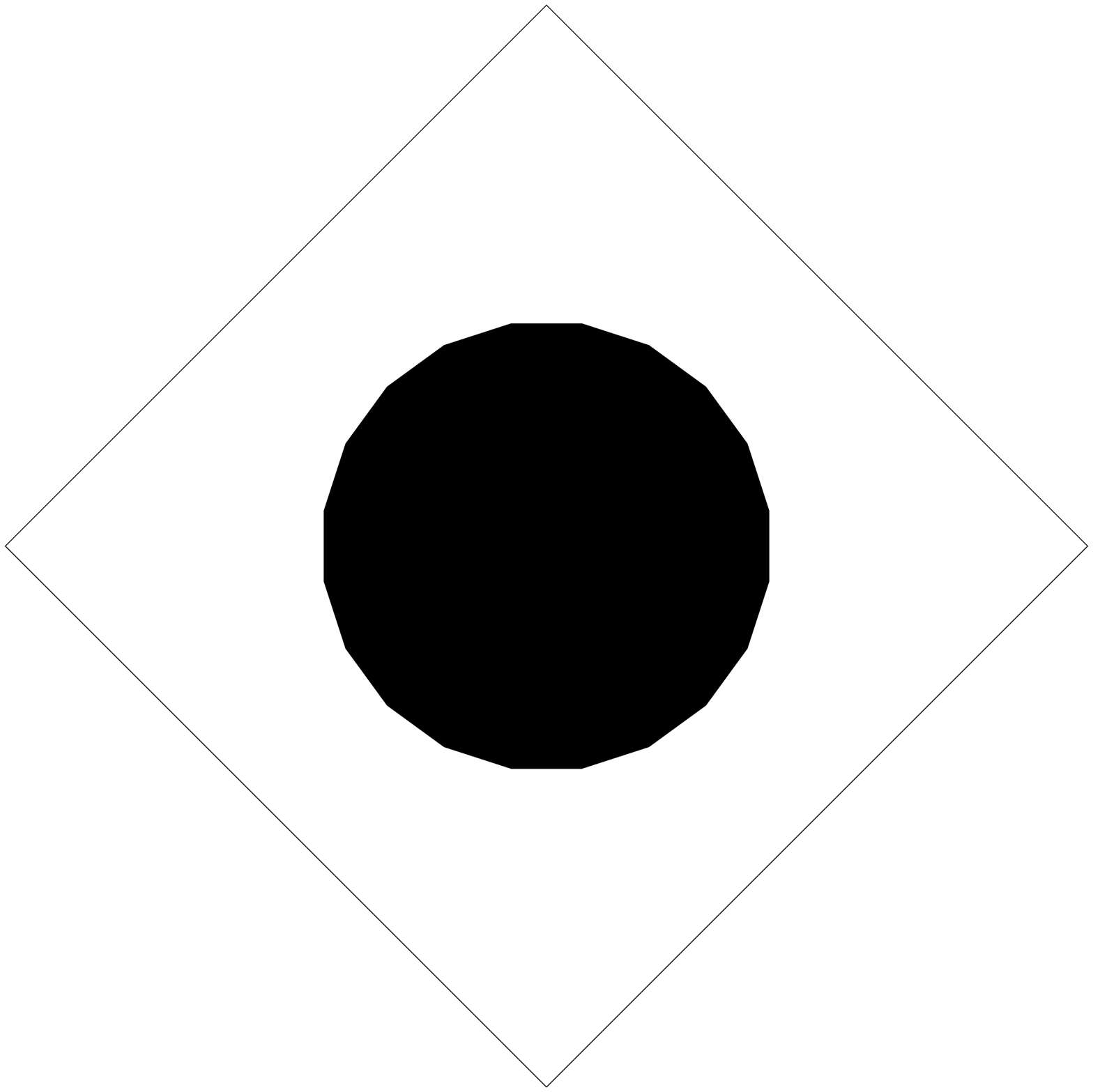}}), and
the Sun ($\odot$). Giant data are for 28 M-stars and 9 K-stars
(+). Dashed and solid lines represent the fits of Eqs. (3) and
(5). Error bars are included for the giant data. For the dwarfs, the
error bars equal the size of the symbols. For illustrative purposes,
the reddening path for $A_{\rm v} = 1$ is indicated on the right. Also
shown are the theoretical curves for dwarfs (solid curve) and for
giants (dashed curve), displaced downward by two units for clarity
(see text for the differences in slope).}
\end{minipage}}
\end{figure*}

Figure 1 compares the visual surface brightness vs. Cousins \vic~
relationships for late-type giants and dwarfs. The giant sample
includes 28 stars from Dumm \& Schild (1998) with spectral types M0 to
M7 ($1.65\,<\,$\vic$\,<\,4.30$) and nine non-variable K-giants from
Dyck et al. (1998). Most of the stars are of luminosity class III with
a few of class II. The $V$ magnitudes and Cousins \vic{} values were
taken from the on-line version of the HIPPARCOS catalogue (entries H5
and H40), the angular diameters are from Dumm \&\ Schild (1998) and
Dyck et al. (1998, and private communication). Most of the giants in
the sample are located within the local bubble where the density of
atomic hydrogen is low (Diamond et al. 1995, Thomas \& Beuermann
1998). Inside the bubble, reddening is usually small and the standard
reddening corrections based on distance and latitude (e.g. Fouqu\'e \&
Gieren 1997) are of limited use. We opted not to apply reddening
corrections, therefore. There may be a problem, however, with
circumstellar absorption in some of the stars of latest spectral type
(see e.g. the comments in Dumm \& Schild 1998). We excluded one star
from the giant sample, $\tau^4$\,Ser, which leaves us with 27
stars. For illustrative purposes, we show in \mbox{Fig. 1} the
reddening path for a standard interstellar absorption of $A_{\rm_ v} =
1$ with $A_{\rm_v} = 2.41\,E_{V-I}$ (Schlegel et al. 1998), but expect
that none of the giants shows this much absorption. Since most of the
M-giants are more or less variable, errors due to the non-simultaneity
of the brightness and angular diameter measurements probably dominate
the observed scatter in \sv. The error bars are, therefore, determined
by quadratically adding the variability amplitude (1/2 of the
difference in fields H50 and H49 in the HIPPARCOS catalogue), a
minimum uncertainty in the $V$-magnitude, taken to be 0.02 mag, and
the error in 5\,log$\phi$. The standard error in \vic{} is taken from
entry H41 of the HIPPARCOS catalogue. A linear fit for the 27 M-giants
yields
\begin{equation}
S_{\rm V, giants}  =  4.71(\pm 0.14) + 1.74(\pm 0.05)\,(V-I_{\rm c})
\end{equation}
which agrees with the fit of Dumm \&\ Schild (1998, their Eq. 5)
within the 1$\sigma$ errors. The scatter in the giant data is a source
of concern for the dwarf-giant comparison. We can not exclude that
this comparison is affected by remaining systematic errors in the
giant data, like selection effects, variability, and circumstellar
reddening, which are difficult to estimate. It is comforting, though,
that Eq. (3) is consistent with the red section of the original
relation of Barnes \&\ Evans (1976) which was given as a function of
Johnson \vrj.  Our result is also consistent with the \sv{} vs. \vk{}
relations for Cepheids, field giants and field supergiants discussed
by Fouqu\'e \& Gieren (1997). Using SIMBAD \mbox{$K$-mag}\-nitudes for
16 of our M-giants, we obtain
\begin{equation}
S_{\rm V, giants}  =  3.82(\pm 0.10) + 1.00(\pm 0.05)\,(V-K)
\end{equation}
which is valid for $3.5\,<\,$\vk$\,<\,7$ and agrees with the red part
of the Fouqu\'e \& Gieren relation (see their Fig. 3).

For K-giants \mbox{(\vic\,$<\,1.65$)}, the relation in Fig. 1
steepens. This change of slope was already noted by Barnes \& Evans
(1976) and Barnes et al. (1977) who also showed that there is no
difference in the surface brightness of giants and dwarfs for stars of
spectral types B--G and found the Sun to fall on the giant relation.
Tying the fit to the Sun at \mbox{\vic\,=\,0.70}, \sv\,=\,4.85, we
obtain \mbox{\sv{} = 2.86 + 2.84(\vic)} for \vic{}$<1.65$. All
Barnes-Evans type relations for giants suggest that the slope of the
relation changes near the transition from K to M-stars. In summary,
the results for giants presented here coincide closely with those of
other authors.

The dwarf data in Fig. 1 include eight young disk (YD) M-dwarfs from
L96 and the YD eclipsing binary YY Gem of spectral type M1+M1. The
angular diameters and effective temperatures \teff{} were derived by
L96 from flux fitting of the Hauschildt et al. (1999) {\it NextGen}
model atmospheres to the observed low-resolution optical/IR
spectra. L96 consider the angular diameters obtained by this approach
to be less fallible to errors in \teff{} than those derived from the
observational bolometric magnitude and \teff. They quote an error of
2\% in the derived angular diameters.
An additional uncertainty in the radii may, however, arise from errors
in the flux fitting procedure.  The integrated luminosities of the
best-fit models quoted by L96 tend to fall below the observed
luminosity, i.e.  \Mbol$'\,=\,-2.5\,{\rm log}(4\pi R^2\sigma T_{\rm
eff}^4/L_{\odot}$) + \Mbols~$>$~\Mbol, with $R$ and \teff{} as derived
by L96 and the $>$ sign implying ``fainter than''.  For the eight L96
YD stars, the flux deficiency averages \mbox{$\sim$0.15 mag} relative
to \Mbol~= \mv~+ $BC_{\rm V}$, with $BC_{\rm V}$ the visual bolometric
correction as given by L96 (their Tab. 6), and \mbox{$\sim$0.10 mag}
relative to \Mbol~= \mk~+ $BC_{\rm K}$, with the bolometric correction
in the $K$-band from Tinney et al. (1993). While it is well known that
bolometric magnitudes are uncertain by as much as 0.10 mag (see
e.g. the discussion in L96 and their Fig. 7), the amount of the flux
deficiency suggests that either the temperatures of L96 are low by
$\sim3$\%, their radii by \mbox{$\sim6$\%}, or both by
correspondingly smaller percentages.  Accepting the radii implies that
the temperatures of the eight YD stars are on the average too low by
90\,K which is well within the temperature errors, whereas accepting
the temperatures leads to angular diameters (not radii which include
the parallax errors) outside the quoted range. We assume, therefore,
that the radii are basically correct and proceed to use them for
calibrating our Barnes-Evans relation for dwarfs. 
The possible $\sim6$\% systematic uncertainty in the angular
diameters corresponds to an uncertainty of 0.13 in \sv{} which may
affect the absolute level of \sv{} but probably not the slope of the
\sv--(\vic) relation.  The L96 YD dwarfs are well observed with an
error in \vic{} less than $\sim 0.05$. Both error bars, in \sv{} and
\vic{}, are of the size of the dwarf data points in Fig. 1.

Finally, the mean angular diameter of the two components of YY Gem was
calculated from their observed radii (Leung \&\ Schneider 1978) and the
HIPPARCOS parallax of 63.2mas. Note that there is no systematic
difference between the surface brightness of YY Gem and that
established for YD dwarfs by the radius scale of L96. 

A linear fit to the data in \mbox{Fig. 1}, i.e. the eight YD
dwarfs from L96 and to YY Gem, yields
\begin{equation}
S_{\rm V, YD\,dwarfs} = 3.99(\pm 0.13) + 1.98(\pm 0.05)\,(V-I_{\rm c})
\end{equation}
which is valid for \vic$\,>\,1.65$. Comparison of Eqs. (3) and
(5) shows that there is a difference in slope of $0.24\pm0.07$ which
is significant at the $3.5\sigma$ level. The absolute levels of the
giant and dwarf fits suggest that the visual surface brightness of
early M-dwarfs is slightly higher than that of M-giants, reaching a
separation of $0.30\pm 0.09$\,mag at spectral type K7/M0 (\vic{}
=1.65).
Since there are no equally reliable radii for K-dwarfs, the extension
to \vic$\,<\,1.65$ is not covered. There is no difference in the
surface brightness of giants and dwarfs for stars of spectral types
B--G (Barnes \& Evans 1976, Barnes et al. 1977), however, which
suggests that the observational dwarf relation in Fig. 1 should
connect to the Sun, that the dwarf/giant difference reaches a maximum
for late K and early M stars, and that the break in the \sv(\vic)
relation near \vic$\,=\,1.65$ is less pronounced for dwarfs than for
giants. This observation suggests the presence of gravity effects in
the surface brightness vs. colour relation.

Figure 1 also compares the observational results with the predictions of
recent theoretical work for late-type giants (Alibert et al. 1999,
Hauschildt et al. in preparation) and late-type dwarfs (BCAH98), both
of solar composition. The \sv(\vic) relationship predicted for
solar-metallicity ZAMS dwarfs with masses from 1.2\,\msun{} down to
0.075\,\msun{} is shown as solid curve (for clarity shifted downward
by two units). The corresponding relationship for giants (dashed
curve) is represented by the post main sequence evolutionary track of
a 12\,\msun{} star, evolved until central carbon ignition. Details of
the calculations can be found in the recent work of Alibert et
al. (1999) on Cepheids which shows a generally good agreement between
models and recent observations in period--magnitude and period--radius
diagrams. We note that tracks from 4\,\msun{} to 12\,\msun{} are very
similar in the \sv{} vs. \vic{} diagram. This is in agreement with the
results of Fouqu\'e \& Gieren (1997), who find that giant and
supergiant surface brightness relations are indistinguishable.
Therefore, the 12\,\msun{} track shown in Fig. 1 is representative of
the \sv(\vic) relationship expected for giant and supergiants, with
gravities log\,$g = 0-3$ and effective temperatures \teff{} =
$3500-7000$\,K.

The differences in the slopes and normalizations of Eqs. (3) and (5)
as well as the different strengths of the break for giants and dwarfs
at the K/M transition (\vic{}\,=\,1.65) agree quantitatively with
those predicted by the stellar models.  For the range in \vic{}
covered here, the theoretical \mbox{\sv(\vic)} relationships for
giants and dwarfs reach a maximum separation of $\Delta$\sv{} = 0.35
mag at \vic\,=\,1.65, very similar to the observed difference of
0.30\,mag from Eqs. (3) and (5). 
Since the giant \sv{} is based on observed angular diameters and
we consider the theoretical prediction of the difference in \sv{}
reliable, the dwarf \sv{} and, hence, the dwarf radius scale can not
be seriously in error either.

For an early M-star of given \mv, a difference in \sv{} for giants and
dwarfs of $\Delta$\sv{} = 0.30 mag corresponds to a difference in
radius of 15\%. I.e., if the giant calibration were used for dwarfs,
the derived radii at \mbox{\vic\,=\,1.65} (spectral type K7) would be
15\% too large.

\begin{figure*}[t]
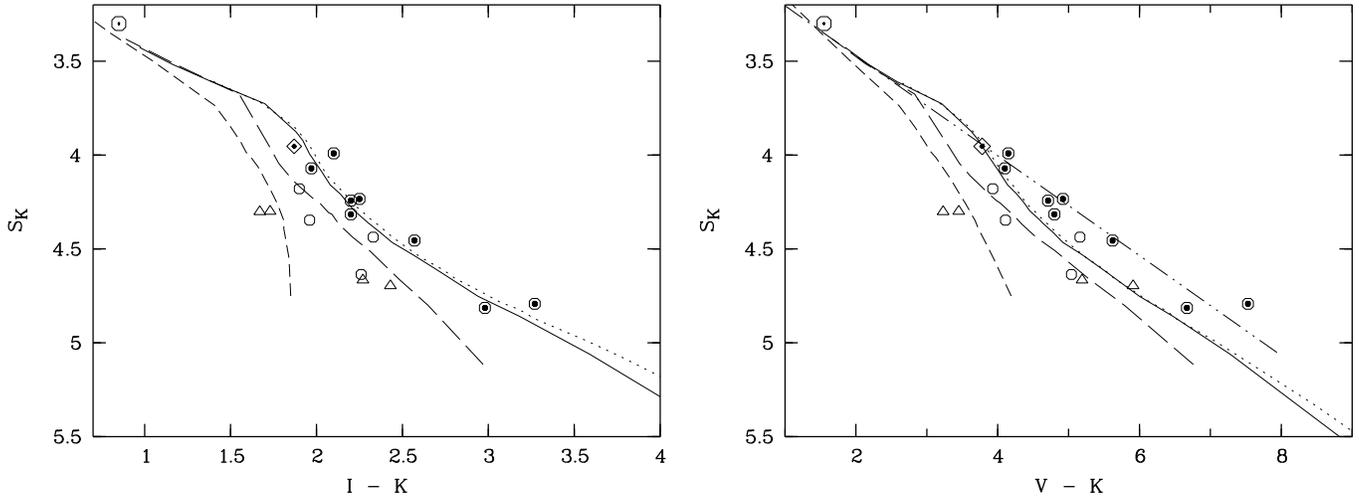
 
\mbox{\includegraphics[width=8.8cm]{beuerman.f2a}}
\hspace*{2mm}
\mbox{\includegraphics[width=8.8cm]{beuerman.f2b}}
\caption[]{\label{mbol_radius} Barnes-Evans type relations \sk{}
vs. \ik{} and \vk{} with the $K$-band magnitudes being on the CIT
system. The data points indicate the L96 dwarfs: YD (\od{} and
\raisebox{-0.5ex}{\pssym{rhombdot.ps}}), OD ($\bigcirc$), and H
($\bigtriangleup$). The curves indicate the theoretical models of
BCAH98 for ZAMS stars with solar metallicity [M/H]= 0 (solid curve),
for solar-metallicity stars aged 0.1 Gyr (dotted), and for stars aged
10 Gyrs with metallicities [M/H]\,$= -0.5$ (long dashes) and [M/H]\,$=
-1.5$ (short dashes).  The dot-dashed line in the right-hand panel (b)
represents the fit of Eq. (6) to the YD data.}
\end{figure*} 

For later spectral types, the difference becomes smaller and even
reverses sign near \vic$\,=\,2.9$, or spectral type M4.5. Models for
giants of still later spectral type (\teff $< 3500$\,K) are not yet
available and the comparison between obervation and theory is
restricted to spectral types earlier than M4.  A quantitative
comparison of the absolute values of \sv{} is limited by the fact that
the theoretical $V$-magnitudes of M-stars calculated with the most
recent models are still somewhat too bright and the colours involving
$V$ too blue (see also Fig. 2b which shows the corresponding effect in
the colour \vk).  This remaining error in the theory is suspected to
be due to uncertainties in the molecular absorption coefficients for
solar-metallicity stars in the visual passband (BCAH98). It causes the
slopes of the \sv(\vic) relations of both M-giants and M-dwarfs at
\vic$~\ga~1.6$ to be too steep, but should, to a first approximation,
not affect the ordinate {\it difference} of the two curves which
measures the gravity effect. Since the error in the theoretical
$V$-magnitudes of late-type stars with near-solar metallicity reaches
about 0.6 mag (see Fig. 3b) the corrected theoretical crossover
between the \sv(\vic) relations of giants and dwarfs occurs at
\vic$\,\simeq 2.9$ rather than \vic$\,= 2.3$, which is just as
observed.

While our results show that a slight gravity effect in the \sv(\vic)
relation is present, the calibrations of Eqs. (3) and (5) are never
further apart than 0.3\,mag. This {\it approximate} equality is the
observational and theoretical basis for Lacy's (1977) approach to
determine the radii of late-type dwarfs from the Barnes-Evans relation
for giants.

\section{Metallicity dependence of \sk{} for M-dwarfs}
 
In addition to gravity, metallicity affects the surface brightness at
a given colour.  Given the steepness of the \sv(\vic) relation,
however, the dependence of the observational data on metallicity is
easily obliterated by errors in \vic. It is more readily detectable in
the infrared $K$-band because \sk{} is a much shallower function of
colour.

The models suggest that the metallicity dependence appears only at
temperatures sufficiently low that molecules are present. For giants
and supergiants with $T_{\rm eff}\,>\,3500$\,K, corresponding to \vic$
< 2$, the models of Alibert et al. (1999) indicate a very small
dependence on metallicity. At these higher temperatures, \sv{} and
\sk{} at given \vic{} or \ik{} vary by less than 0.05 mag for
metallicities between [M/H] = 0 and --0.7. For dwarfs, the main
metallicity effects appear for \vic$\,\ga\,1.8$ (BCAH98), a regime
which is largely unexplored for giants.

In Figure 2ab, we compare the observed values of \sk{} vs. \ik{} and
\vk{} for the 16 dwarfs of L96 supplemented by YY Gem and the Sun with
the predictions of the models of BCAH98 for main-sequence dwarfs. The
theoretical curves refer to ZAMS models with solar metallicity
[M/H]\,$= 0$ (solid curves) and to models for stars aged 10 Gyr with
metallicities $-0.5$ (long dashes) and $-1.5$ (short dashes).
Pre-main-sequence stars of solar composition aged $10^8$ years (dotted
curves) and ZAMS stars of the same colour agree closely in surface
brightness, except for very late spectral types where the $10^8$-yrs
isochrone is sufficiently far from the ZAMS for gravity effects to
become apparent. For these pre-main-sequence stars \sk{} is enhanced
and differs from ZAMS dwarfs in the same way as is evident for
late-type giants in Fig. 1.

\begin{table}[t]
\caption[ ]{Model values of the surface brightness \sk{} in the K-band
as functions of \ik{} and \vk{} for ZAMS stars with solar metallicity
and stars aged 10 Gyr with 1/3 solar metallicity (from BCAH98). The
$K$-band magnitudes are on the CIT system.
}
\begin{flushleft}
\begin{tabular}{lrccrccc}
\noalign{\smallskip} \hline \noalign{\smallskip}
Mass & \multicolumn{3}{c}{ZAMS, [M/H] = 0} & \multicolumn{4}{c}{10 Gyrs, 
[M/H]$ = -0.5$}\\[0.3ex] 
\msun & \mk & \ik & \sk & \mk & \ik & \vk & \sk  \\ [0.3ex]
\noalign{\smallskip} \hline \noalign{\smallskip}
1.00 & 3.49 & 0.97 & 3.43 &  &  &  &  \\
0.80 & 3.98 & 1.30 & 3.57 & 3.62 & 0.89 & 1.67 & 3.38 \\  
0.70 & 4.49 & 1.65 & 3.70 & 4.26 & 1.17 & 2.13 & 3.51 \\  
0.60 & 5.08 & 1.92 & 3.89 & 4.82 & 1.55 & 2.83 & 3.68 \\  
0.50 & 5.74 & 2.06 & 4.11 & 5.56 & 1.73 & 3.32 & 3.96 \\  
0.45 & 6.04 & 2.11 & 4.16 & 5.92 & 1.78 & 3.46 & 4.04 \\  
0.40 & 6.32 & 2.16 & 4.21 & 6.25 & 1.84 & 3.61 & 4.11 \\  
0.35 & 6.59 & 2.20 & 4.26 & 6.55 & 1.88 & 3.72 & 4.15 \\  
0.30 & 6.88 & 2.23 & 4.29 & 6.83 & 1.92 & 3.81 & 4.18 \\  
0.25 & 7.24 & 2.29 & 4.33 & 7.20 & 1.99 & 3.97 & 4.24 \\  
0.20 & 7.71 & 2.37 & 4.41 & 7.67 & 2.09 & 4.21 & 4.32 \\  
0.175& 7.99 & 2.44 & 4.45 & 7.96 & 2.13 & 4.34 & 4.37 \\  
0.150& 8.33 & 2.55 & 4.53 & 8.30 & 2.21 & 4.54 & 4.44 \\  
0.130& 8.66 & 2.70 & 4.61 & 8.63 & 2.31 & 4.81 & 4.51 \\  
0.110& 9.09 & 2.95 & 4.74 & 9.10 & 2.48 & 5.29 & 4.66 \\  
0.100& 9.39 & 3.17 & 4.85 & 9.44 & 2.65 & 5.79 & 4.80 \\  
0.090& 9.81 & 3.58 & 5.06 &10.00 & 2.97 & 6.76 & 5.12 \\  
0.080&10.57 & 4.40 & 5.54 &11.50 & 3.92 & 9.48 & 6.27 \\    
\noalign{\medskip}\hline
\end{tabular}
\end{flushleft}
\end{table}

Figure 2a demonstrates that the main-sequence models reproduce the
observed level of \sk, its variation with \ik, and the spread with
metallicity exceedingly well. The agreement between observation and
theory is less good for \sk{} vs. \vk{} in Fig. 2b because, as noted
above, the theoretical colours of solar-metallicity M-stars with
\vk$~\ga~4$ which involve $V$ are too blue (by about half a
magnitude). This uncertainty in $V$ is much smaller for dwarfs of
lower metallicity. For the purpose of determining radii via Eq. (2),
we provide the surface brightness values of the BCAH98 models in
Tab. 1, except for \sk(\vk) of stars with solar metallicity for which
we approximate the data for the eight L96 YD dwarfs, YY Gem, and the
Sun by the {\it linear} relation
\begin{equation}
S_{\rm K, YD\,dwarfs} = 2.95(\pm 0.08) + 0.266(\pm 0.017)\,(V-K).
\end{equation}
The limited statistics of the L96 sample does not warrant a
higher-order fit, but the real \sk(\vk) relation for solar-metallicity
stars will certainly show some structure caused by molecule formation
and the onset of convection in the optically thin layers of the
atmosphere, as does the \sk(\ik) relation at \ik~ $\simeq\,1.9$.

\section{The $R$(\mk) relation of [M/H]$\,\simeq 0$ ZAMS dwarfs}

For stars of given age and metallicity, theory provides the radius as
a function of absolute magnitude, e.g. \mk. In the present context
this follows from the fact that the models yield \vk{} as a function
of \mk{} which transforms Eq. (6) into \sk(\mk) and Eq. (2) into
$R$(\mk).

Fig. 3a shows the observed radii of YY Gem (mean component) and the
eight YD, four old disk (OD), and four \mbox{halo (H)} M-dwarfs of L96
along with the BCAH98 model radii for solar-composition ZAMS stars
(solid curve, see also Fig. 1 of \mbox{Paper I}). There is no obvious
difference between the radii of YD and OD stars in this rather
restricted sample. The two faint H stars show the expected smaller
radii. This is consistent with the small metallicity dependence of the
$R$(\mk) BCAH98 models which can approximately be expressed by
$\Delta$log$R\,\simeq\,-0.03$ per 1 dex reduction of [M/H] relative to
solar (BCAH98). On the average, the L96 radii of the YD/OD stars
exceed the [M/H] = 0 ZAMS model radii by 2\% (Paper I) which is within
the systematic uncertainties of the L96 radii. These model radii
(solid curve) can be represented reasonably well by a third-order
polynomial in \mk{} which we adjust slightly, by $\Delta {\rm log} R =
+0.009$, to nominally fit the L96 radii (dot-dashed curve in Fig. 3a)
\begin{equation}
{\rm log}\frac{R}{R_{\odot}}=-0.022+0.1294M_{\rm K}-0.04464M_{\rm
K}^2+0.002237M_{\rm K}^3.
\end{equation}  
We accept Eq. (7) as representative of ZAMS stars with near-solar or
slightly reduced metallicities (kinematic classes YD and OD). Note
that radii for stars fainter than \mk$~\simeq\,10$ are still uncertain
because dust formation is not accounted for in the BCAH98 models

\begin{figure}[tb]
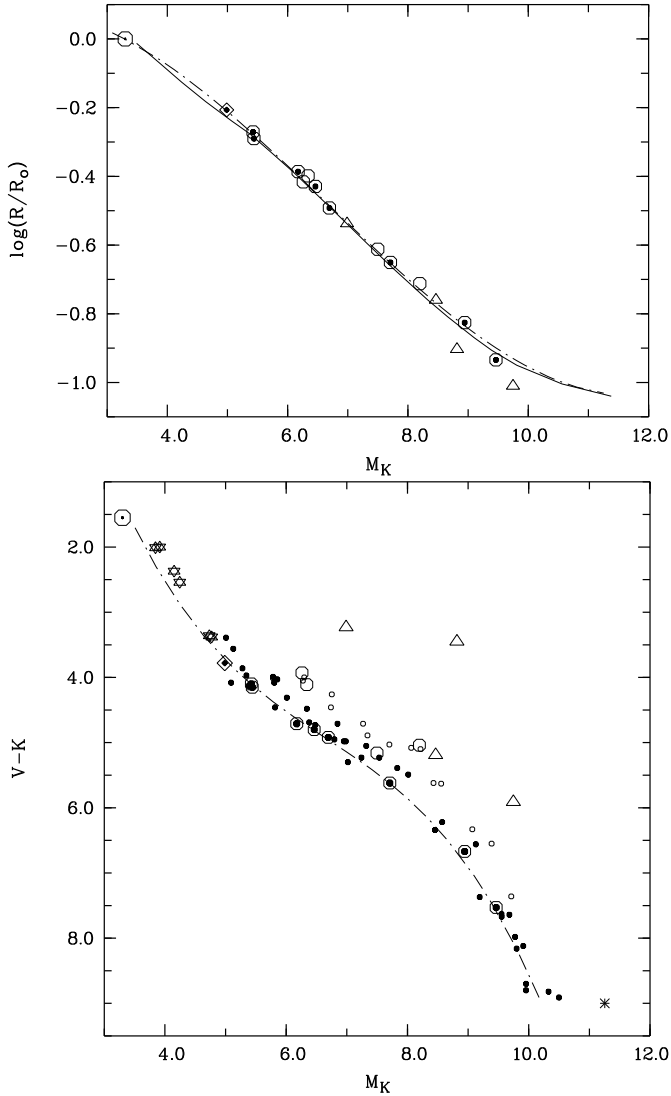

\mbox{\includegraphics[width=8.8cm]{beuerman.f3a}}
\mbox{\includegraphics[width=8.8cm]{beuerman.f3b}}
\caption[]{{\it (a) Top panel: }Observationally determined radii of YY
Gem and the L96 dwarfs (symbols as in Fig. 2) compared with the
solar-metallicity ZAMS model radii of BCAH98 (solid curve). Also shown
is the polynomial representation of the model curve by Eq. 7
(dot-dashed curve). (b) {\it Bottom panel: } Colour-magnitude diagram
\vk{} vs. \mk. The stars from Henry \& McCarthy (1993) are shown as
small solid and open circles (see text), the stars from Reid \& Gizis
(1997) as stars and GD165B as an asterisk. The dot-dashed curve
represents the location of ZAMS stars with near-solar metallicity
according to Eq. 8.}
\end{figure}

\begin{table*}[t]
\caption[ ]{Radii of YD and OD dwarfs from L96.  The columns indicate
(1) the name, (2) the spectral type, (3) the kinematic population
class, (4) the parallax in mas from L96, (5) \vk, (6) \ik, (7) the
absolute K-magnitude, in case of binaries the mean of the two
components, \mbox{(8) the logarithm} of the radius in units of \rsun{}
as derived observationally by L96, (9) the radius obtained from \vk{}
in column 5 with Eqs. 2 and 6 for the [M/H]$\simeq 0$ stars and with
\sk(\vk) in Tab. 1 for the [M/H]$\simeq -0.5$ stars, (10) the radius
derived from \ik{} in column 5 with \sk(\ik) in Tab. 1 and Eq. 2,
(11) the radius from Eq. 7, and (12) the radius given by Lacy (1977)
corrected to the parallax in column 3.}
\begin{center}
\begin{tabular}{llccccccccccc}
\noalign{\smallskip} \hline \noalign{\smallskip}
 (1) & (2) & (3) & (4) & (5) & (6) & (7) && (8) & (9) & (10) & (11) & (12) \\[0.3ex]
Name & $Sp$ & Pop & $\pi$ & $V-K$ & $I-K$ & $M_{\rm K}$ &&
\multicolumn{5}{c}{log\,($R$\,/\rsun) from}\\ [0.3ex]
& & & mas & & & && L96 & \sk(\vk) & \sk(\ik) & $R$(\mk) & Lacy (1977) \\[0.3ex]
\noalign{\smallskip}\hline\noalign{\medskip}
\multicolumn{13}{l}{\it (1) Young disk M-dwarfs with [M/H]\,$\simeq\,0$}\\
[0.3ex]
Gl65$\overline{AB}^{1)}$  
      &M6--&YD&373.8&6.67&2.95&8.94&&$-0.83\pm0.03$&--0.85&--0.84&--0.84&--0.81\\
Gl195A&M2: &YD& 76.4&4.10&1.97&5.43&&$-0.27\pm0.03$&--0.28&--0.28&--0.28&\\ 
Gl206$\overline{AB}^{1)}$  
      &M3.5&YD& 73.6&4.92&2.25&6.69&&$-0.49\pm0.04$&--0.49&--0.48&--0.49&\\ 
Gl268$\overline{AB}^{1)}$  
      &M4.5&YD&165.1&5.62&2.57&7.71&&$-0.65\pm0.01$&--0.66&--0.63&--0.65&--0.65\\ 
Gl388 &M3  &YD&205.5&4.71&2.20&6.17&&$-0.39\pm0.02$&--0.40&--0.38&--0.40&--0.30\\ 
Gl494A$^{2)}$ &M1.5&YD& 92.5&4.15&2.10&5.44&&$-0.29\pm0.04$&--0.28&--0.26&--0.28&\\ 
Gl896A&M3.5&YD&150.1&4.80&2.20&6.46&&$-0.43\pm0.03$&--0.45&--0.44&--0.45&\\ 
GJ1111&M6.5&YD&275.8&7.53&3.27&9.46&&$-0.93\pm0.02$&--0.90&--0.92&--0.90&--0.97\\
[0.8ex]
\multicolumn{13}{l}{\it (2) Old disk M-dwarfs with [M/H]\,$\simeq\,-0.5$}\\
[0.3ex]
Gl213&M4&O/H&168.1&5.16&2.33&7.50&&$-0.61\pm0.02$&--0.58&--0.59&--0.62&--0.51\\ 
Gl411&M2&OD &394.4&4.11&1.96&6.34&&$-0.40\pm0.02$&--0.41&--0.43&--0.43&--0.33\\ 
Gl699&M4&O/H&546.0&5.04&2.26&8.20&&$-0.71\pm0.03$&--0.72&--0.75&--0.73&--0.68\\ 
Gl908&M1&OD &174.6&3.93&1.90&6.26&&$-0.42\pm0.02$&--0.41&--0.42&--0.41&--0.34\\ 
\noalign{\medskip}\hline\noalign{\smallskip}
\end{tabular}
$^{1)}~\overline{AB}$ refers to the mean of the two binary components.
$^{2)}$~Following L96, the contribution by the faint secondary is neglected.  
\end{center}
\end{table*}

\begin{table*}[hbt]
\caption[] {Radii of single or presumed single field M-dwarfs from the
list of Henry \& McCarthy (1993). The columns indicate (1) the name,
(2) the spectral type, (3) the kinematic population class if available
(4) the parallax, (5) \vk, (6) \ik, (7) the absolute K-magnitude, (8)
the logarithm of the radius in units of \rsun{} as derived from Eqs. 2
and 6, (9) same as derived from the theoretical \sk(\ik) relation of
BCAH98 for [M/H] = 0 in Fig. 2a, (10) same as derived from Eq. 7, and
(11) the radius as given by Lacy (1977), adjusted to the parallax in
column 4.}
\begin{center}
\begin{tabular}{llcrccrccccc}
\noalign{\smallskip} \hline \noalign{\smallskip} 
(1) & (2) & (3) &\multicolumn{1}{c}{(4)} & (5) & (6) & (7) && (8) & 
(9) & (10) & (11) \\[0.3ex] 
Name&$Sp$&Pop&\multicolumn{1}{c}{$\pi$}&\vk&\ik&\mk&&\multicolumn{4}{c}
{log\,($R$\,/\rsun) from}\\ [0.3ex] 
& & &\multicolumn{1}{c}{mas}& & & && \sk(\vk) & \sk(\ik)  & $R$(\mk) & 
Lacy (1977)\\
[0.3ex]
\noalign{\smallskip}\hline\noalign{\medskip} 
\multicolumn{12}{l}{\it (1) Brighter M-dwarfs, assumed to have
[M/H]\,$\simeq\,0$}\\[0.6ex]
 Gl68   & K1  &    &138.9&2.01&1.15&  3.84&&--0.07&--0.07&--0.06&--0.07 \\
 Gl105A & K3  &    &125.6&2.37&1.30&  4.15&&--0.12&--0.12&--0.09&--0.07 \\
 Gl105B & M4- & O  &138.7&5.05&2.26&  7.32&&--0.61&--0.60&--0.59&--0.54 \\
 Gl109  & M3+ & Y  &125.6&4.69&2.24&  6.38&&--0.44&--0.42&--0.43&--0.37 \\
 Gl166A & K1  & O  &198.2&2.00&1.12&  3.91&&--0.09&--0.08&--0.07&--0.08 \\
 Gl166C & M4.5& O  &198.2&5.23&2.36&  7.53&&--0.64&--0.63&--0.62& \\
 Gl205  & M1.5& O  &175.5&4.08&2.01&  5.09&&--0.21&--0.21&--0.23&--0.11 \\
 Gl229  & M1.5& Y  &173.2&3.97&1.96&  5.34&&--0.27&--0.27&--0.26&--0.16 \\
 Gl251  & M3  & Y/O&173.7&4.73&2.23&  6.48&&--0.46&--0.44&--0.45&--0.43 \\
 Gl273  & M3.5& O  &263.3&4.98&2.28&  6.98&&--0.54&--0.53&--0.53&--0.45 \\
 Gl300  & M4+ & Y  &170.0&5.39&2.49&  7.83&&--0.69&--0.67&--0.67& \\
 Gl338A & M0  & Y  &162.0&3.56&1.82&  5.13&&--0.25&--0.26&--0.23&--0.16 \\
 Gl380  & K7  & Y/O&205.2&3.38&1.76&  4.77&&--0.19&--0.20&--0.18&--0.06 \\
 Gl393  & M2  & Y/O&136.2&4.31&2.07&  6.01&&--0.39&--0.38&--0.37&--0.30 \\
 Gl402  & M4  & Y/O&145.9&5.23&2.44&  7.24&&--0.58&--0.55&--0.58& \\
 Gl406  & M6  & O  &419.1&7.37&3.31&  9.19&&--0.86&--0.86&--0.87&--0.83 \\
 Gl408  & M3  & Y  &144.6&4.48&2.09&  6.34&&--0.44&--0.44&--0.43&--0.35 \\
 Gl447  & M4+ & O  &299.6&5.49&2.51&  8.01&&--0.72&--0.70&--0.70&--0.68 \\
 Gl450  & M2  & O  &107.8&4.08&2.00&  5.80&&--0.36&--0.35&--0.34& \\
 Gl514  & M1  & O  &139.5&3.99&1.98&  5.78&&--0.36&--0.35&--0.33&--0.24 \\
 Gl555  & M4  & Y/O&159.0&5.30&2.44&  7.02&&--0.53&--0.51&--0.54&--0.44 \\
 Gl570A & K4  &    &169.3&2.54&1.20&  4.24&&--0.13&--0.14&--0.11&--0.06 \\
 Gl581  & M3.5& Y/O&158.2&4.71&2.20&  6.85&&--0.53&--0.52&--0.51&--0.44 \\
 Gl628  & M3.5& Y  &234.5&4.98&2.30&  6.95&&--0.54&--0.52&--0.53&--0.47 \\
 Gl644C & M7  & O  &154.0&7.98&3.42&  9.78&&--0.95&--0.96&--0.93& \\
 Gl673  & K7  & O  &129.5&3.36&1.75&  4.73&&--0.18&--0.19&--0.17&--0.04 \\
 Gl701  & M2  & O  &126.6&4.03&1.97&  5.85&&--0.37&--0.37&--0.35&--0.28 \\
 Gl752A & M3  & O  &170.3&4.46&2.14&  5.82&&--0.34&--0.32&--0.34&--0.24 \\
 Gl752B & M8  & O  &170.3&8.70&4.00&  9.96&&--0.94&--0.93&--0.95& \\
 Gl809  & M1  & O  &134.2&3.86&1.87&  5.28&&--0.26&--0.28&--0.25&--0.17 \\
 Gl873  & M3.5& Y/O&198.1&4.95&2.26&  6.79&&--0.51&--0.50&--0.50&--0.43 \\
 Gl880  & M2  & O  &148.5&4.13&2.03&  5.39&&--0.27&--0.26&--0.27&--0.18 \\
 Gl884  & M0- & O  &128.1&3.39&1.75&  5.01&&--0.23&--0.25&--0.21&--0.10 \\
 Gl905  & M5.5& O  &316.8&6.34&2.89&  8.45&&--0.77&--0.75&--0.77&--0.72 \\
 GJ1156 & M5  & Y  &101.5&6.22&2.76&  8.57&&--0.80&--0.78&--0.78& \\
 GJ1245B& M5.5&    &220.2&6.56&2.83&  9.16&&--0.89&--0.89&--0.86& \\
 LHS191 & M6.5&    & 59.2&7.63&3.26&  9.55&&--0.92&--0.93&--0.91& \\
 LHS292 & M6.5& O  &220.9&7.64&3.24&  9.68&&--0.94&--0.96&--0.92& \\
 LHS2065& M9  &    &117.3&8.82&4.46& 10.33&&--1.01:&--0.95:&--0.98:& \\
 LHS2397a&M8  & O  & 68.7&8.80&4.18&  9.95&&--0.94&--0.91&--0.95& \\
 LHS2471& M7  & O  & 70.9&7.67&3.39&  9.55&&--0.92&--0.92&--0.91& \\
 LHS2924& M9  & O  & 92.4&8.91&4.54& 10.50&&--1.04:&--0.98:&--1.00:& \\
 LHS2930& M7  & Y/O&103.8&8.16&3.59&  9.80&&--0.94&--0.95&--0.94& \\
 LHS3003& M7  &    &157.0&8.12&3.60&  9.91&&--0.96&--0.97&--0.95& \\
 GD165B & L   &    & 26.6&    &5.03& 11.25&&      &--1.02:&--1.03:& \\[0.8ex]
\noalign{\smallskip}\hline\noalign{\smallskip}
\end{tabular}
\end{center}
\vspace*{-6mm}
\end{table*}

\begin{table*}[hbt]
\hspace*{15mm}{\bf Table 3 continued}
\begin{center}
\begin{tabular}{llcrccrccccc}
\noalign{\smallskip} \hline \noalign{\smallskip} 
(1) & (2) & (3) &\multicolumn{1}{c}{(4)} & (5) & (6) & (7) && (8) & (9) & (10) & (11) \\
[0.3ex] 
Name&$Sp$&Pop&\multicolumn{1}{c}{$\pi$}&\vk&\ik&\mk&&\multicolumn{4}{c}{log\,($R$\,/\rsun) from}\\ 
[0.3ex] 
& & &\multicolumn{1}{c}{mas}& & & && \sk(\vk) & \sk(\ik)  & $R$(\mk) & Lacy (1977)\\
[0.3ex]
\noalign{\smallskip}\hline\noalign{\medskip} 
\multicolumn{12}{l}{\it (2) Fainter M-dwarfs, assumed to have
[M/H]\,$\simeq\,-0.5$}\\[0.6ex]
 Gl015A & M1.5& O  &281.2&4.05&1.91&  6.28&&--0.40&--0.42&--0.41&--0.38 \\
 Gl015B & M4- & O  &281.2&5.10&2.28&  8.22&&--0.72&--0.75&--0.73&--0.70 \\
 Gl054.1& M4.5& O  &267.4&5.63&2.48&  8.56&&--0.76&--0.78&--0.78& \\
 G3-33  & M4.5& O  &224.8&5.62&2.50&  8.43&&--0.73&--0.75&--0.76&--0.61 \\
 Gl412A & M1  & O  &203.0&4.00&1.98&  6.30&&--0.41&--0.41&--0.42&--0.32 \\
 Gl412B & M6  & O  &203.0&6.55&2.78&  9.39&&--0.87&--0.89&--0.89& \\
 Gl445  & M3.5& O  &191.5&4.89&2.25&  7.34&&--0.56&--0.58&--0.59&--0.49 \\
 Gl526  & M1.5& O  &183.7&4.01&1.97&  5.78&&--0.31&--0.31&--0.33&--0.27 \\
 Gl625  & M2  & Y  &152.2&4.26&2.05&  6.75&&--0.48&--0.49&--0.49& \\
 Gl643  & M4  & O  &155.7&5.03&2.30&  7.70&&--0.63&--0.64&--0.65&--0.63 \\
 Gl725A & M3  & Y/O&288.1&4.46&2.00&  6.74&&--0.47&--0.50&--0.49& \\
 Gl725B & M3.5& Y/O&288.1&4.71&2.16&  7.27&&--0.56&--0.58&--0.58& \\
 Gl729  & M4- & Y/O&342.3&5.08&2.30&  8.06&&--0.69&--0.71&--0.71&--0.87 \\
 GJ1002 & M5.5& O  &212.8&6.33&2.73&  9.07&&--0.82&--0.84&--0.85& \\
 LHS523 & M6.5& O  & 91.7&7.36&3.10&  9.71&&--0.88&--0.89&--0.93& \\
\noalign{\smallskip}\hline
\end{tabular}
\end{center}
\end{table*}

Stellar radii may be estimated either from a Barnes-Evans type
relation as Eq. (6) together with Eq. (2) or directly from the
$R$(\mk) relation in Eq. (7). The former depends weakly on gravity but
distinctly on metallicity, while the latter depends weakly on
metallicity and strongly on gravity and, therefore, requires knowledge
of the evolutionary status.

The different metallicity dependencies of the two approaches arise
because reference is taken to a colour in one case and directly to the
abolute magnitude in the other. The difference in the approaches
becomes more obvious when combining Eqs. (2), (6), and (7) to yield
the colour-magnitude relation \vk{} vs. \mk{} for ZAMS stars of
near-solar metallicity
\begin{equation}
V-K=-11.50+6.215\,M_{\rm K}-0.8423\,M_{\rm K}^2+0.04221\,M_{\rm K}^3.
\end{equation}  
(Fig. 3b, dot-dashed curve). Eq. (8) closely fits the L96 YD stars
(encircled dots) which is as expected because Eqs. (6) and (7) fit
these stars, too. Although Eq. (7) is approximately valid also for OD
stars with slightly reduced metallicity, Eq. (6) and Eq.  (8) are
not. At the {\it same \mk}, stars of lower metallicity (open points,
triangles in Fig. 3b) are bluer. They have nearly unchanged \sk{} and
radii, however, because the metallicity dependence of the colour in
Fig. 3b and the metallicity dependence of \sk(\vk) in Fig. 2b
compensate approximately. This implies that application of the
Barnes-Evans relations in \mbox{Fig. 2} requires knowledge of the
metallicity of the respective stars.

\section{Radii of main-sequence M-dwarfs}

In this section, we test our results on the 12 YD/OD stars of L96
which serve as calibrators and then apply them to a sample of 60
single or presumed single YD/OD main-sequence stars. The
colour-magnitude diagram of these stars is shown in
\mbox{Fig. 3b}. Six K-stars are from Reid \& Gizis (1997) (stars)
and 53 M-stars from Henry \& McCarthy (1993) (small solid and open
circles). The L-star GD165B represents the transition to the
brown-dwarf regime (asterisk).

We consider the calibrator stars first. In Tab. 2, we compare the L96
radii with those obtained from the Barnes-Evans type relations
\sk(\vk) and \sk(\ik) and from \mk:
\begin{description}
\item Column 9: Radii of YD or [M/H]$\,\simeq 0$ stars derived from
the observed \vk{} with Eqs. (2) and (6). Radii of OD stars obtained
correspondingly, but with the theoretical \sk(\vk) relation for
\mbox{[M/H]$ \,= -0.5$} of Tab. 1 instead of \mbox{Eq. (6)}.
\item Column 10: Radii obtained from the observed \ik{} with the
theoretical \sk(\ik) relations of Tab. 1 for the \mbox{[M/H] = 0} or the
[M/H]\,=\,$-0.5$ stars.
\item Column 11: Radii derived from \mk{} with Eq. (7) without regard
of metallicity, assuming the stars to be on the ZAMS.
\end{description}
The radii in column 11 and the YD radii in column 9 agree with the L96
radii within 5\% or $\Delta$log$R = 0.02$ which reflects the goodness
of the fits. The small differences between the radii in columns 10 and
8, on the other hand, demonstrate the close agreement between theory and
observation. 

Application of the Barnes-Evans type relations to the complete sample
of 60 K/M dwarfs requires at least a rough estimate of their
metallicity. For this purpose, we divide the sample into a {\it
brighter} and a {\it fainter} subsample, containing stars within~
$\sim 1$ mag of the bright limit in \mk{} (small solid circles in
Fig. 3b) and stars within $\sim 1-2$ mag from the bright limit (small
open circles), respectively. This subdivision can be interpreted in
terms of different metallicity if age is not an influencing
factor. Assuming all stars to be close to the ZAMS, the two subsamples
correspond to stars of near-solar metallicity and a metallicity
reduced by $\Delta[M/H] \simeq 0.5$, respectively (e.g. L96). In
deriving the radii, we proceed as above for the L96 stars.  Table 3
provides the observed properties and the derived radii. The
restriction to single stars is important in order to avoid the larger
radii falsely assigned to unrecognized binaries.

\begin{figure}[t]
\mbox{\includegraphics[width=8.8cm]{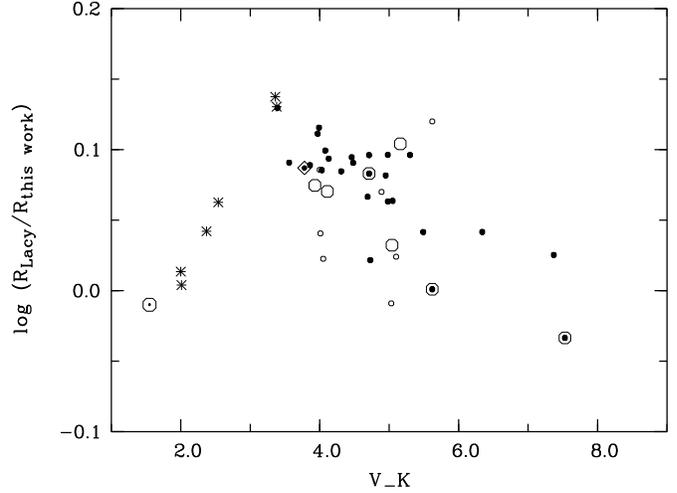}}
\caption[]{Differences in log($R$/\rsun) between the radii given by
Lacy (1977) and here: (i) L96 YD/OD stars (Tab. 2, column 8, large
symbols as in Fig. 2) and (ii) stars in Tab. 3, radii derived from
\sk(\vk) (column 8, small symbols as in Fig. 3 ).}
\end{figure}

Column 12 of Tab. 2 and column 11 of Tab. 3 list the radii given by
Lacy (1977) adjusted to the parallaxes used here.  Compared with
Lacy's results, our radii are smaller by up to $\sim 25$\%. Fig.4
shows that there is a systematic trend for the difference between
Lacy's radii and those determined by L96 or, for the addtional stars,
from our Barnes-Evans type relations \sk(\vk). Very similar pictures
obtain for the radii derived from the Barnes-Evans type relation
\sk(\ik) (Tab. 3, column 9) or directly from \mk{} (Tab. 3, column 10,
both not shown). The deviation of Lacy's from our radii assumes a
maximum at spectral type K7 and vanishes for early K and for late M
dwarfs.  Much of this is due to the different surface brightness
calibrations for giants (used by Lacy) and dwarfs (used here) which
reach a maximum separation at \vic{} $\simeq 1.6$, \vk{} $\simeq 3.4$, or
spectral class K7 (Fig. 1). The remainder is  due to differences in the
individual giant relations used by Lacy (Barnes \& Evans 1976) and by
us (Eq. 3 and Dumm \& Schild 1997).
Lacy discussed a deviation of similar magnitude and colour-dependence
between the surface brightness of giants and the theoretical ZAMS
dwarf models of Copeland et al. (1970). He interpreted it as being
entirely due to inadequacies of the models. We now know that (i) the
surface brightness of mid-K to mid-M dwarfs is, in fact, higher than
that of giants of the same colour and (ii) the recent dwarf models
(BCAH98) predict somewhat larger radii than the early models which
reduces the discrepancy noted by Lacy (1977).

Finally, we discuss the systematic errors in our radius calibration
which is tied to the observational results of L96 and to the
theoretical predictions of BCAH98. Our radius scale for stars of
near-solar metallicity is based on the results of L96 which may still
be in error by some $\sim 6$\% or 0.03 in log\,$R$ (radii too small,
see Sect. 2). The difference of this radius scale to the BCAH
[M/H]\,=\,0 model is $\sim 2$\%. Clemens et al. (1998), on the other
hand, quote radii for stars later than M2 {\it at given mass} which
are larger than those predicted by BCAH98 by $\sim$20\%. Part of this
discrepancy may be due to remaining uncertainties in the bolometric
corrections (Sect. 2) and part to the transformations used by
them. Clemens et al. adopt the \Mbol{} scale and the temperature scale
of L96 to deduce radii which are higher by $\sim 7$\% than those
quoted by L96. They employ the {\it mean observational} \mv($M$)
relation of Henry \& McCarthy (1993) in order to convert absolute
visual magnitudes \mv{} to masses $M$. For $M\,<\,0.2$\,\msun, this
relation presently relies on 10 stars only (Henry et al. 1999). It
shows substantial scatter which may be caused by a spread in ages and
metallicity and/or still by the inclusion of erroneous masses. The
masses of the YD binary Wolf\,424 ( $0.143\pm0.011$ and
$0.131\pm0.010$\,\msun) were recently re-determined with the {\it HST}
Fine Guidance Sensors (Torres et al. 1999) and agree perfectly with
the predictions of BCAH98. We expect that a better definition of the
mass--radius relation will become available soon (Henry et
al. 1999) and allow the remaining discrepancies to be resolved.

\section{Conclusions}
 
We conclude that the effects of gravity on the visual surface
brightness of M-giants and M-dwarfs as well as the effects of
metallicity on the surface brightness of M-dwarfs are discernible in
the data and agree quantitatively with the predictions of recent
stellar models (Alibert et al. 1999, Hauschildt et al. in preparation,
BCAH98). The surface brightness values of \mbox{giants} and dwarfs
agree for spectral types earlier than $\sim$K2 and later than $\sim$M5
and reach a maximum separation at spectral type K7.  Although small,
this difference must be taken into account when estimating dwarf radii
by the surface brightness method.

We present improved Barnes-Evans type relations which allow to
determine the radii of late-type giants and dwarfs of known distances
with a remaining systematic uncertainty of $\sim 6$\%. Our calibration
is based on the M-dwarf radii determined by L96 from fits of the {\it
NextGen} model atmospheres of Hauschildt et al. (1999) to the observed
spectra of these stars. While these radii are not purely
observational, the internal consistency of atmosphere calculations,
stellar models, and observation has reached a high degree of
excellency. We conclude that the differences between observed and
predicted surface brightness for stars on the lower main sequence have
largely been resolved.

\acknowledgements{We thank Mel Dyck for sending us his photometry of
giants and Boris G\"ansicke, Frederick Hessman, and Klaus
\mbox{Reinsch} for many discussions and the referee H. Schild for
helpful comments. This research has made use of the on-line version of
the HIPPARCOS catalogue.  IB thanks the Universit\"ats-Sternwarte,
G\"ottingen, for hospitality and the APAPE (PROCOPE contract 97151)
for travel support.  The work of PH was supported in part by NASA ATP
grant NAG 5-3018, LTSA grant NAG 5-3619 and NSF grant AST-9720804 to
the University of Georgia. Some of the calculations presented in this
paper were performed on the IBM SP2 and SGI Origin 2000 of the UGA
UCNS, at the San Diego Supercomputer Center (SDSC) and at the National
Center for Supercomputing Applications (NCSA), with support from the
National Science Foundation, and at the NERSC with support from the
DoE. We thank all these institutions for a generous allocation of
computer time.}

\end{document}